\documentclass[preprint]{JHEP3} 

\JHEPspecialurl{http://jhep.sissa.it/JOURNAL/JHEP3.tar.gz}

\usepackage{epsfig,multicol,bbm,cite}

\newcommand\fverb{\setbox\pippobox=\hbox\bgroup\verb}
\newcommand\fverbdo{\egroup\medskip\noindent%
			\fbox{\unhbox\pippobox}\ }
\newcommand\fverbit{\egroup\item[\fbox{\unhbox\pippobox}]}
\newbox\pippobox

\newcommand{\Fig}[1]{figure~\ref{#1}}

\newcommand{\Sec}[1]{section~\ref{#1}}
\newcommand{\Tab}[1]{table~\ref{#1}}

\newcommand{\Rs}{\ensuremath{R_S}}
\newcommand{\Rl}{\ensuremath{R_L}}
\newcommand{\Rsg}{\ensuremath{R_{S,\gamma}}}

\def\ifm#1{\relax\ifmmode#1\else$#1$\fi}  
    \def\vv{\vphantom{H^I_I}}
\def\x{\ifm{\times}}   \def\pt#1,#2,{\ifm{#1\x10^{#2}}}
\def\up#1{\ifm{^{#1}}}     \def\plm{\ifm{\,\pm}\,}
   \def\deg{\ifm{^\circ}}  
\def\gam{\ifm{\gamma}}  \def\eps{\ifm{\epsilon}}  \def\to{\ifm{\rightarrow}} 
\def\kl{\ifm{K_L}}   \def\ks{\ifm{K_S}}
\def\eiii{\ifm{\pi^\pm e^\mp\nu}}  

\def\muiii{\ifm{\pi^\pm \mu^\mp\nu}}   

\def\pio{\ifm{\pi^0\pi^0}} 
\def\po{\ifm{\pi^0}}
\def\pic{\ifm{\pi^+\pi^-}}  

\def\rmk{\rm\kern.5mm }

\def\eq#1{eq.~(\ref{#1})}
\def\GAML{\ifm{\Gamma_L}}
\def\GAMS{\ifm{\Gamma_S}}
\def\ko{\ifm{K^0\vphantom{\overline K {}^0}}}
\def\kob{\ifm{\overline K {}^0}}

\newcommand{\beq}{\begin{equation}}
\newcommand{\eeq}{\end{equation}}
\newcommand{\ba}{\begin{array}}
\newcommand{\ea}{\end{array}}
\newcommand{\beqa}{\begin{eqnarray}}
\newcommand{\eeqa}{\end{eqnarray}}
\newcommand{\no}{\nonumber}
\newcommand{\dis}{\displaystyle}
\newcommand{\cO}{\mathcal{O}}
\newcommand{\cA}{\mathcal{A}}
\def\Gno{\ifm{\Gamma_{\ko}}}
\def\Gnb{\ifm{\Gamma_{\kob}}}
\def\Mno{\ifm{m_{\ko}}}
\def\Mnb{\ifm{m_{\kob}}}
\def\BR{{\rm BR}}
\def\Re{{\rm Re}}
\def\Im{{\rm Im}}

\title{\mathversion{bold} Determination  of $CP$ and $CPT$ violation parameters
 in the neutral kaon system using the Bell-Steinberger relation and data from the KLOE experiment}
\author{The KLOE collaboration:\\
F.~Ambrosino,$^a$
A.~Antonelli,$^b$
M.~Antonelli,$^b$
C.~Bacci,$^c$
P.~Beltrame,$^d$
G.~Bencivenni,$^b$
S.~Bertolucci,$^b$
C.~Bini,$^e$
C.~Bloise,$^b$
S.~Bocchetta,$^c$
V.~Bocci,$^e$
F.~Bossi,$^b$
D.~Bowring,$^{b,f}$
P.~Branchini,$^c$
R.~Caloi,$^e$
P.~Campana,$^b$
G.~Capon,$^b$
T.~Capussela,$^a$
F.~Ceradini,$^c$
S.~Chi,$^b$
G.~Chiefari,$^a$
P.~Ciambrone,$^b$
S.~Conetti,$^f$
E.~De~Lucia,$^b$
A.~De~Santis,$^e$
P.~De~Simone,$^b$
G.~De~Zorzi,$^e$
S.~Dell'Agnello,$^b$
A.~Denig,$^d$
A.~Di~Domenico,$^e$
C.~Di~Donato,$a$
S.~Di~Falco,$^g$
B.~Di~Micco,$^c$
A.~Doria,$^a$
M.~Dreucci,$^b$
G.~Felici,$^b$
A.~Ferrari,$^b$
M.~L.~Ferrer,$^b$
G.~Finocchiaro,$^b$
S.~Fiore,$^e$
C.~Forti,$^b$
P.~Franzini,$^e$
C.~Gatti,$^b$
P.~Gauzzi,$^e$
S.~Giovannella,$^b$
E.~Gorini,$^h$
E.~Graziani,$^c$
M.~Incagli,$^g$
W.~Kluge,$^d$
V.~Kulikov,$^i$
F.~Lacava,$^e$
G.~Lanfranchi,$^b$
J.~Lee-Franzini,$^{b,l}$
D.~Leone,$^d$
M.~Martini,$^b$
P.~Massarotti,$^a$
W.~Mei,$^b$
S.~Meola,$^a$
S.~Miscetti,$^b$
M.~Moulson,$^b$
S.~M\"uller,$^b$
F.~Murtas,$^b$
M.~Napolitano,$^a$
F.~Nguyen,$^c$
M.~Palutan,$^b$
E.~Pasqualucci,$^e$
A.~Passeri,$^c$
V.~Patera,$^{b,m}$
F.~Perfetto,$^a$
L.~Pontecorvo,$^e$
M.~Primavera,$^h$
P.~Santangelo,$^b$
E.~Santovetti,$^n$
G.~Saracino,$^a$
B.~Sciascia,$^b$
A.~Sciubba,$^{b,m}$
F.~Scuri,$^g$
I.~Sfiligoi,$^b$
A.~Sibidanov,$^{b,k}$
T.~Spadaro,$^b$
M.~Testa,$^e$
L.~Tortora,$^c$
P.~Valente,$^e$
B.~Valeriani,$^d$
G.~Venanzoni,$^b$
S.~Veneziano,$^e$
A.~Ventura,$^h$
R.Versaci,$^b$
G.~Xu,$^{b,j}$}
\author{and\\
G.~D'Ambrosio,$^a$ G.~Isidori~$^b$
\\
\setcounter{page}{0}
\thispagestyle{empty}
\\
\llap{$^a$}Dipartimento di Scienze Fisiche dell'Universit\`a  ``Federico II'' e Sezione INFN, Napoli, Italy\\
\llap{$^b$}Laboratori Nazionali di Frascati dell'INFN, Frascati, Italy\\
\llap{$^c$}Dipartimento di Fisica dell'Universit\`a ``Roma Tre'' e Sezione INFN, Roma, Italy\\
\llap{$^d$}Institut f\"ur Experimentelle Kernphysik, Universit\"at Karlsruhe, Germany\\
\llap{$^e$}Dipartimento di Fisica dell'Universit\`a ``La Sapienza'' e Sezione INFN, Roma, Italy\\
\llap{$^f$}Physics Department, University of Virginia, Charlottesville, VA, USA\\
\llap{$^g$}Dipartimento di Fisica dell'Universit\`a e Sezione INFN, Pisa, Italy\\
\llap{$^h$}Dipartimento di Fisica dell'Universit\`a e Sezione INFN, Lecce, Italy\\
\llap{$^i$}Institute for Theoretical and Experimental Physics, Moscow, Russia\\
\llap{$^j$}Institute of High Energy Physics of Academia Sinica,  Beijing, China}

\author{\\
\llap{$^k$}Budker Institute of Nuclear Physics, Novosibirsk, Russia\\
\llap{$^l$}Physics Department, State University of New York at Stony Brook, USA\\
\llap{$^m$}Dipartimento di Energetica dell'Universit\`a ``La Sapienza'', Roma, Italy\\
\llap{$^n$}Dipartimento di Fisica dell'Universit\`a ``Tor Vergata'' e Sezione INFN, Roma, Italy\\
}

\received{\today} 		
\revised{\today}
\accepted{\today}		

\abstract{
  We present an improved determination of the $CP$ and $CPT$ violation
  parameters $\Re(\epsilon)$ and $\Im(\delta)$  based on the
  unitarity condition (Bell-Steinberger relation) and on recent results
  from the KLOE experiment. We find $ \Re(\epsilon) = (159.6 \pm 1.3)\times 10^{-5}$ and $\Im(\delta) = (0.4 \pm 2.1)\times 10^{-5}$, consistent with no
$CPT$ violation.
}


\begin{document} 

\section{Introduction}
\label{sec:intro}

The three discrete symmetries of quantum mechanics, charge conjugation ($C$), parity ($P$) and time reversal ($T$), are known to be violated in nature, both singly and in bilinear combinations. Only $CPT$ appears to be an exact symmetry of nature. Exact $CPT$ invariance holds in quantum field theory, which assumes Lorentz invariance (flat space), locality and unitarity~\cite{CPT}. 
Testing the validity of $CPT$ invariance therefore probes the most fundamental assumptions of our present understanding of particles and their interactions.
These hypotheses are likely to be violated 
at very high energy scales, where quantum effects of the gravitational 
interaction cannot be ignored~\cite{CPTV}. 
On the other hand, since we still lack a consistent theory of quantum gravity,
it is hard to predict at which level violation of $CPT$ invariance might become experimentally observable. 

The neutral kaon system offers unique possibilities for the study of $CPT$ invariance. 
From the requirement of unitarity, Bell and Steinberger have derived a relation, the so-called 
Bell-Steinberger relation (BSR)~\cite{BS:paper}. 
The BSR relates a possible violation of $CPT$ invariance ($m_{\ko} \not= m_{\kob}$ and/or $\Gno\not=\Gamma_{\kob}$) 
in the time-evolution of the $\ko$--$\kob$ system to the observable $CP$-violating interference of 
$K_L$ and $K_S$ decays into the same 
final state $f$. Strictly speaking, evidence of $CPT$ violation found via the BSR 
could just be a failure of the unitarity assumption. 
However, unitarity is also one of the main hypotheses of the $CPT$ theorem; 
thus the BSR allows a test of the basic assumptions of quantum field theories. 
 
In this work we use recent results from the KLOE experiment to improve the determination of the phenomenological $CP$- and $CPT$-violating parameters  $\Re(\epsilon)$ 
and $\Im(\delta)$ by means of the BSR. 
 Our analysis benefits in particular from three new measurements: i) the branching ratio 
\BR(\kl\to\pic )~\cite{KLOE:bpp},
 which is relevant to the determination of  $\Re(\epsilon)$; ii) the new upper limit on 
\BR(\ks\to\pio\po )~\cite{KLOE:BKS3pi0}, which is necessary to improve the accuracy on  
$\Im(\delta)$; and iii) 
the measurement of the semileptonic charge asymmetry $A_S$~\cite{KLOE:BKSpienu}, which
allows, for the first time, the complete determination of the contribution from semileptonic 
decay channels {\it without assuming unitarity}. 

 A determination  of $\Re(\epsilon)$ and $\Im(\delta)$ using the BSR was performed by 
CPLEAR~\cite{CPLEAR:bs} in 1999. In the analysis of Ref.~\citen{CPLEAR:bs}, some of the parameters
of the semileptonic channels were evaluated together with $\Re(\epsilon)$ and $\Im(\delta)$
from a combined fit to the time-dependent semileptonic asymmetries
{\it imposing the constraint of the} BSR.
A recent update of the determination of  $\Re(\epsilon)$ and $\Im(\delta)$ is given in 
Ref.~\citen{NA48:BRS3pi0}. In this latter analysis, however, some of the results of the CPLEAR fit \cite{CPLEAR:bs}, {\it which used the unitarity constraint}, have been used as input again to the BSR. It is not clear whether this fact {\it was accounted for} in \cite{NA48:BRS3pi0}.

Our presentation is organized as follows. 
In \Sec{sec:th} we outline the meaning of the BSR and the approach
used to maximize the sensitivity obtainable from the presently available data. 
In \Sec{sec:alpha_i} we examine experimental inputs, including a re-analysis of the semileptonic channels, and obtain a best estimate for errors and correlations. 
The extraction of $\Re(\epsilon)$ 
and $\Im(\delta)$ is discussed in \Sec{sec:resu}.

\section{Theoretical framework}
\label{sec:th}

Within the Wigner-Weisskopf approximation, the time evolution 
of the neutral kaon system is described by~\cite{WW}
\beq
i {\partial \over \partial  t} \Psi(t) = H\Psi(t) = \left(M-{i\over 2}\Gamma\right)\Psi(t)~,
\label{LOY}
\eeq
where $M$ and  $\Gamma$ are 2\x2 time-independent 
Hermitian matrices and $\Psi(t)$ is a two-component 
state vector in the $\ko$--$\kob$ space. 
Denoting by $m_{ij}$ and $\Gamma_{ij}$ the elements 
of $M$ and  $\Gamma$ in the $\ko$--$\kob$ basis, 
$CPT$ invariance implies 
\beq 
m_{11} = m_{22} \quad ({\rm or}~\Mno=\Mnb) \qquad{\rm and}\qquad 
\Gamma_{11} = \Gamma_{22} \quad ({\rm or}~\Gno=\Gnb)~.
\eeq
The eigenstates of \eq{LOY} can be written as
\beqa
    K_{S,L} &=& 
        \dis\frac{1}{\sqrt{2\left( 1 + |\epsilon_{S,L}|^2 \right)}}  
                  \left( \left(1 + \epsilon_{S,L}\right)\ko \pm
                  \left( 1 - \epsilon_{S,L} \right) \kob \right),\\
\epsilon_{S,L} &=&  \frac{
    -i \Im\left( m_{12}\right) - 
            \frac{1}{2} \Im\left(\Gamma_{12}\right) \pm
    \frac{1}{2} \left( \Mnb - \Mno -\frac{i}{2} 
    \left( \Gnb -\Gno\right) \right)
                    }{
    m_L - m_S +i(\Gamma_S - \Gamma_L)/2  
                    } \no\\
& \equiv & \epsilon \pm \delta~,
\label{epsLS}
\eeqa
such that $\delta=0$ in the limit of exact $CPT$ invariance.

Unitarity allows us to express the four elements of $\Gamma$ in terms 
of appropriate combinations of the kaon decay amplitudes $\cA_i$:
\beq
 \Gamma_{ij} = \sum_f \cA_i (f) \cA_j(f)^*,\qquad i,j=1,2=\ko,\kob,
\eeq
where the sum runs over all the accessible final states. Using this decomposition 
in \eq{epsLS} leads to  the BSR: a link between 
$\Re(\epsilon)$, $\Im(\delta)$, and the physical kaon decay amplitudes.
In particular, without any expansion in the $CPT$-conserving parameters 
and neglecting only $\cO(\epsilon)$ corrections to the coefficient of the 
$CPT$-violating parameter $\delta$, we find 
\beq
     \ \kern-2em\left({{\GAMS+\GAML}\over{\GAMS-\GAML}}+i\tan\phi_{\rm SW}\right)
      \left(\frac{\Re(\epsilon)}{1+|\epsilon|^2 }-i\Im(\delta) \right) = 
      {1\over{\GAMS-\GAML}} \sum_f \cA_L(f) \cA^*_S(f),
\label{eq:b-s}
\eeq
where $\phi_{\rm SW} = \arctan\left(\vv 2 (m_L -m_S)/(\GAMS-\GAML)\right)$.
We stress that, in contrast to similar expressions which can be found in the 
literature, \eq{eq:b-s} is exact and phase-convention independent
in the exact $CPT$ limit: any evidence for a non-vanishing $\Im(\delta)$ 
resulting from this relation can only be attributed to violations of: 
i) $CPT$ invariance; ii) unitarity; iii) the time independence 
of $M$ and $\Gamma$ in \eq{LOY}.

The advantage of the neutral kaon system is that only a few decay modes 
give significant contributions to the r.h.s. in \eq{eq:b-s}. Only the  $\pi\pi (\gamma)$, $\pi\pi\pi $ and $\pi\ell\nu$ modes turn out to be relevant up to the $10^{-7}$ level\footnote{Note that all quantum numbers 
of a chosen final state $f$ have to be equal between \ks\ and \kl\ decays in order to allow for 
interference between the two amplitudes in the r.h.s. of \eq{eq:b-s}.}. 
  
The products of the corresponding decay amplitudes are conveniently 
expressed in terms of the  $\alpha_i$ parameters defined below.

\subsection{Two-pion modes}
\label{sec:ppg}

For two-pion states, we define the ratios $\alpha_i$ as:
\beq
\alpha_i \equiv {1\over\GAMS}\:\langle
 \cA_L(i) \cA^*_S(i) \rangle = \eta_i~ \BR(\ks \to i)~, \qquad  
i=\pio~,~\pic(\gamma),
\label{eq:alfa2pi} 
\eeq
where $\pic(\gamma)$ denotes the inclusive sum over bremsstrahlung photons, 
and $\langle\ldots \rangle$ indicates the appropriate phase-space integrals. 
The $\eta_i$ parameters in \eq{eq:alfa2pi} are the usual amplitude ratios: 
$\eta_i=\cA_L(i)/\cA_S(i)$. 

The contributions from $\pi^+\pi^-\gamma$ direct-emission (DE)
amplitudes not included in $\alpha_{\pi^+\pi^-(\gamma)}$ 
are collected together in the term
\beq
\alpha_{\pi\pi\gamma_{\rm DE}}  = \alpha_{\pi\pi\gamma_{\rm E1-S}} + \alpha_{\pi\pi\gamma_{\rm E1-L}} +
\alpha_{\pi\pi\gamma_{{\rm DE} \times {\rm DE}}}~, 
\label{eq:alfappg1}
\eeq
where 
\beqa 
\alpha_{\pi\pi\gamma_{\rm E1-S}} &+& \alpha_{\pi\pi\gamma_{\rm E1-L}}~ = \nonumber \\ 
&=& {1\over\GAMS} \left( \vv\langle \cA_L(\pi\pi\gamma)  \cA^*_S(\pi\pi\gamma_{\rm E1}) \rangle +  \langle 
\cA_L(\pi\pi\gamma_{\rm E1})   \cA^*_S(\pi\pi\gamma)\rangle \right)\nonumber \\ 
&=& \Delta B(\ks\to\pi\pi\gamma_{\rm DE})~\eta_{+-} + (\eta_{+-\gamma}-\eta_{+-})
~ \BR(\ks \to \pi\pi\gamma).
\label{eq:alfappg2}
\eeqa
Here $\cA_{L,S}(\pi\pi\gamma)$ and $\cA_{L,S}(\pi\pi\gamma_{\rm E1})$
denote the leading  bremsstrahlung and the 
electric-dipole DE amplitudes, respectively. Their interference cannot be trivially neglected.  
$\BR(\ks \break\to  \pi\pi\gamma)$ indicates the branching fraction for decays with the emission
of a real photon with minimum energy equal to the cut used in the 
corresponding $\eta_{+-\gamma}$  measurement.
$\Delta B(\ks\to\pi\pi\gamma_{\rm DE}) = \BR(\ks\to\pi\pi\gamma)^{\rm exp} - 
\BR(\ks\to\pi\pi\gamma)^{\rm th-IB}$ is 
the DE contribution to the BR, obtained subtracting the computed bremsstrahlung spectrum
from the measured spectrum (see appendix~\ref{sec:appendix}).

We have generically denoted by $\alpha_{\pi\pi\gamma_{{\rm DE} \times {\rm DE}}}$ 
the contribution arising from the product of two DE amplitudes 
(electric or magnetic). Given the strong
experimental suppression of DE amplitudes,
this term turns out to be safely 
negligible being of $\cO(10^{-8})$ or less~\cite{DI}.  

\subsection{Three-pion modes}

For three-pion states we define
\beq
   \alpha_i \equiv {1\over\GAMS} \langle \cA_L(i) \cA^*_S(i) \rangle
={\tau_{\ks}\over \tau_{\kl}} \eta^*_i ~ \BR(\kl \to i), 
\qquad  i = 3\po~,~\po \pi^+\pi^-(\gamma).
\label{eq:alfa3pi}
\eeq
Note that in this case the amplitudes are not necessarily constant over
the phase space. As a result, the $\eta_i$ appearing in \eq{eq:alfa3pi}
should be interpreted as appropriate Dalitz-plot averages.
In particular, the $\pi^+\pi^-\po$ final state is not a 
$CP$ eigenstate. For decays to $\pi^+\pi^-\po$ the $\eta$ parameter can be expressed as  
\beqa
\eta_{+-0} &=& \frac{ \left\langle\vv\cA_L^*(\pi^+\pi^-\pi^0,\ CP=+1)\cA_S(\pi^+\pi^-\pi^0,\ CP=+1)\right \rangle }{\langle \left| \cA_L(\pi^+\pi^-\pi^0) \right|^2 \rangle } + \no \\
&+& \frac{ \left \langle \vv\cA_L^*(\pi^+\pi^-\pi^0,\ CP=-1)\cA_S(\pi^+\pi^-\pi^0,\ CP=-1) \right \rangle }
{  \langle \left| \cA_L(\pi^+\pi^-\pi^0)  \right|^2 \rangle }.
\eeqa
The experimental bounds on $\eta_{+-0}$ reported by CPLEAR \cite{CPLEAR} 
correspond to this average, when neglecting the contribution of $\cA_L(\pi^+\pi^-\pi^0,\ CP=+1)$. This is indeed
a good approximation, because the \kl\ decay amplitude to a $CP=+1$ \pic\po\  state is suppressed both by $CP$ 
violation and a centrifugal barrier.
Given the poor direct experimental 
information on $\eta_{000}$, in the neutral case it turns out to be 
more convenient to set a bound on $| \alpha_{\pio\po}|$ by means of
 the relation
\beq
|\alpha_{\pio\po}|^2={\tau_{\ks} \over\tau_{\kl}}\:\BR(\kl \to 3\po)\x\BR(\ks\to 3\po )~. 
\label{eq:a000}
\eeq 
This relation is based on the well-justified 
assumption that the \kl\ (\ks) decays to 3\po\ are dominated by a single $CP$ conserving 
(violating) amplitude with the same behaviour over phase space~\cite{DI}.

\subsection{Semileptonic modes}

In the case of semileptonic channels,  
the standard decomposition is~\cite{handbook}
\beqa
  \cA(\ko \to \pi^-l^+\nu)       &=&  \cA_0(1-y)~, \nonumber \\
  \cA(\ko \to \pi^+l^-\nu)       &=&  \cA_0^* (1+ y^*) (x_+ - x_-)^*~,  \nonumber \\
  \cA(\kob \to \pi^+l^-\nu) &=&  \cA_0^* (1+ y^*)~, \nonumber \\
  \cA(\kob \to \pi^-l^+\nu) &=&  \cA_0(1-y) (x_+ + x_-)~,
\label{ampkl}
\eeqa
where $x_+$ ($x_-$) describes the violation of the $\Delta S =\Delta Q$ rule in $CPT$ conserving (violating) decay amplitudes,
 and $y$ parametrizes $CPT$ violation for $\Delta S=\Delta Q$ transitions.
 Assuming lepton universality, and expanding to the first non-trivial 
order in the small $CP$- and $CPT$-violating parameters, one obtains 
\begin{eqnarray}
& \sum_{\pi\ell\nu} \langle \cA_L(\pi\ell\nu) \cA^*_S(\pi\ell\nu) \rangle = 
2 \Gamma(\kl \to \pi\ell\nu)\left(\vv\Re(\epsilon)-\Re(y)-i ( \Im(x_+)+\Im(\delta))\right) &
 \no  \\
& \qquad =  
2 \Gamma(\kl \to \pi\ell\nu)\left(\vv(A_S+A_L)/4-i(\Im(x_+)+\Im(\delta))\right)~. &
\label{eq:kl3}
\end{eqnarray}
The dependence of $\Re(y)$ has been eliminated by taking advantage of the 
relation $\Re(\epsilon)-\Re(y) = (A_S+A_L)/4 $~\cite{handbook},
where $A_{L,S}$ are the observable  semileptonic charge asymmetries. 
The parameter $\Im(x_+)$ can be measured using the appropriate time-dependent 
decay distributions~\cite{CPLEAR:semi}, while $\Im(\delta)$ is one of the two
outputs of the BSR. In order to get rid of the explicit 
$\Im(\delta)$ dependence, it is convenient to define 
\begin{eqnarray}
\label{eq:alfasl}
\alpha_{\pi\ell\nu}
&\equiv& {1\over\GAMS} \sum_{\pi\ell\nu}  \langle \cA_L(\pi\ell\nu) \cA^*_S(\pi\ell\nu) \rangle 
+2i{\tau_{\ks}\over \tau_{\kl}}\:\BR(\kl \to \pi\ell\nu) \Im(\delta) \nonumber \\
&=& 2{\tau_{\ks}\over \tau_{\kl}}\: \BR(\kl \to \pi\ell\nu)\left(\vv(A_S+A_L)/4-i \Im(x_+)\right)~. 
\end{eqnarray}

\subsection{Determination of $\Re(\epsilon)$ and $\Im(\delta)$}
The $\alpha_i$ parameters defined in eqs.~(\ref{eq:alfa2pi}), (\ref{eq:alfappg1}), (\ref{eq:alfa3pi}), 
and (\ref{eq:alfasl}) can be determined (or bounded)
in terms of measurable quantities. Taking into account these definitions
(in particular the non-standard expression for $\alpha_{\pi\ell\nu}$), 
the solution to the unitarity relation in \eq{eq:b-s} is
\beq
\pmatrix{\dis{\Re(\eps)\over1+|\epsilon|^2}\cr\noalign{\vglue-2mm}\cr\Im(\delta)\cr}=
{1\over N}
\pmatrix{1+\kappa(1-2b) & (1-\kappa) \tan\phi_{\rm SW}\cr\noalign{\vglue-2mm}\cr
(1-\kappa) \tan\phi_{\rm SW}  & -(1+\kappa)\cr}
\pmatrix{\Sigma_i \Re(\alpha_i)\cr\noalign{\vglue-4mm}\cr\Sigma_i \Im(\alpha_i)\cr}, 
\label{eq:b-s_sol} 
\eeq
where $\kappa=\tau_{\ks}/\tau_{\kl}$,  $b=\BR(\kl\to \pi \ell \nu)$, and 
\beq
N = (1+\kappa)^2+(1-\kappa)^2  \tan^2\phi_{\rm SW} -2\,b\,\kappa(1+\kappa)~.
\eeq

\section{Experimental input to the $\alpha$ parameters}
\label{sec:alpha_i}

 The experimental inputs needed for the determination of the $\alpha_i$
 are the \kl\ and \ks\ branching ratios,
 the amplitude ratios $\eta_i$, and the \kl\ and \ks\
 lifetimes. All experimental inputs used in the determination of
 the decay amplitudes are summarized in \Tab{tab:inputdata}.  
\begin{table}[!ht]
\begin{center}
\begin{tabular}{c|c|c}
                   &        Value               &  Source \\ \hline
$\tau_{\ks}$       & 0.08958\plm 0.00005 ns     &  PDG~\cite{PDG}  \\ 
$\tau_{\kl}$       & 50.84\plm0.23 ns           &  KLOE average    \\ 
$m_L - m_S$        &\pt(5.290 \pm 0.016),9, s\up{-1}&  PDG~\cite{PDG} \\ 
\BR(\ks\to\pic)    & 0.69186\plm0.00051         &  KLOE average  \\ 
\BR(\ks\to\pio)    & 0.30687\plm 0.00051        &  KLOE average  \\ 
\BR$(\ks\to\pi e\nu)$& \pt(11.77\pm 0.15),-4,   &  KLOE~\cite{KLOE:BKSpienu} \\ 
\BR(\kl\to\pic)    & \pt(1.933 \pm 0.021),-3,   &  KLOE average  \\ 
\BR(\kl\to\pio)    & \pt(0.848 \pm 0.010),-3,   &  KLOE average  \\ 
$ \phi_{+-}$       &(43.4\plm 0.7)\deg          &  PDG~\cite{PDG}   \\ 
$ \phi_{00}$       &(43.7\plm 0.8)\deg          &  PDG~\cite{PDG}  \\ 
$\Rsg$ ($E_\gam >20$MeV)&\pt(0.710\pm0.016),-2, &  E731~\cite{BRKSppg} \\ 
$\Rsg^{\rm th-IB}$ ($E_\gamma >20$MeV)
                   &\pt(0.700 \pm 0.001),-2,    &  KLOE MC~\cite{gattirad} \\  
$|\eta_{+-\gamma}|$&\pt(2.359 \pm 0.074),-3,      &  E773~\cite{etappg}  \\ 
$ \phi_{+-\gamma}$ &(43.8\plm 4.0)\deg             &  E773~\cite{etappg}  \\ 
\BR(\kl\to\pic\po) &0.1262\plm 0.0011           &  KLOE average \\ 
$ \eta_{+-0}$      &\pt\left(\vv(-2\pm7)+i(-2\pm9)\right),-3,& CPLEAR~\cite{CPLEAR}  \\ 
\BR(\kl\to 3\po)   &0.1996\plm 0.0021           &  KLOE average   \\ 
\BR(\ks\to 3\po)   &\pt<1.5,-7, at 95\%\ CL     &  KLOE~\cite{KLOE:BKS3pi0} \\ 
$ \phi_{000}$      & uniform from 0 to 2$\pi$                 &                \\ 
\BR$(\kl\to\pi\ell\nu)$&0.6709\plm 0.0017       &  KLOE average  \\ 
$ A_L + A_S $      &\pt(0.5 \pm 1.0),-2,        &  $K_{\ell3}$ average   \\ 
$ \Im(x_+)   $      &\pt(0.8 \pm 0.7),-2,        &  $K_{\ell3}$ average   \\ 
\end{tabular}
\vskip 0.3cm
\caption{Input values to the Bell-Steinberger relation. Results from \cite{PDG} are those evaluated without assuming $CPT$ invariance.
 The KLOE average and the $K_{\ell3}$ average are described in Sec.~3 and Sec.~3.3, respectively.} 
\label{tab:inputdata}
\end{center}
\end{table}
For the determination of the neutral kaon decay rates and the \kl\ lifetime we
combine the measurements listed below.
\begin{enumerate}
\item The absolute \kl\ \BR's from KLOE~\cite{KLOE:brl};
\item The KLOE \kl\ lifetime ~\cite{KLOE:KLlife};
\item The KLOE ratio  $\BR(\kl\to\pic)/\BR(\kl\to\pi\mu\nu)$~\cite{KLOE:bpp}.
This result is inclusive of final
  state radiation, therefore the DE contribution in the process \kl \to \pic\
   is subtracted using the result from Ref.~\cite{BRKSppg};
\item The precise KLOE determination of $\Rs=\BR(\ks\to\pic)/\BR(\ks\to\pio)$
~\cite{KLOE:BrS}, and  the value of $\Rs/\Rl$   from the
world average of $Re(\epsilon'/\epsilon)$~\cite{PDG}, with $\Rl=\BR(\kl\to\pic)/\BR(\kl\to\pio)$. These
 measurements are used to determine the value of \BR(\kl\to\pio). 
\end{enumerate}
 The combination of all of the above listed measurements, referred to as the KLOE average, 
accounts for all correlation effects, and
is obtained by renormalizing the sum of \kl\ branching ratios to $1-\BR(\kl\to\gamma\gamma)$.
This procedure yields very small corrections to the published BR values~\cite{KLOE:brl}.
 The results with errors and correlation coefficients are given in \Tab{tab:KLOEbr}.
\begin{table}[hbt]
\begin{center}
\renewcommand{\arraystretch}{1.1} 
\renewcommand{\tabcolsep}{0.35pc} 
\begin{tabular}{@{}l|c|ccccccccc}
                     &   Value                      & \multicolumn{8}{|c}{Correlation coefficients} \\ \hline
\BR(\kl \to \eiii)   & 0.4009(15)                   &    1  &       &       &       &      &      &  & \\
\BR(\kl \to \muiii)  & 0.2700(14)                   & -0.31 &  1    &       &       &      &      &  & \\
\BR(\kl \to \pic\po) & 0.1262(11)                   & -0.01 & -0.14 &  1    &       &      &      &  & \\
\BR(\kl \to \pio\po) & 0.1996(20)                   & -0.54 & -0.41 & -0.47 &  1    &      &      &  & \\
\BR(\kl \to \pic)    & 1.933(21)\x10\up{-3} & -0.15 &  0.50 & -0.06 & -0.23 & 1    &      &  & \\   
\BR(\kl \to \pio)    & 8.48(10)\x10\up{-4}  & -0.14 &  0.49 & -0.06 & -0.23 & 0.97 & 1    &  &  \\  
$\tau_{\kl}$~(ns)    & 50.84(23)                    &  0.16  & 0.21 & -0.26 & -0.13 & 0.07 & 0.06 & 1 & \\
\Rs                  & 2.2549(54)                   &  0.00 &  0.00 & 0.00  &  0.00 & 0.00 &-0.21 & 0.00& 1 \\
\end{tabular}\\[2pt]
\caption{Values, errors, and correlation coefficients for all parameters included in the KLOE average: the
dominant \kl\ \BR's, \kl\ lifetime and \Rs.}
\label{tab:KLOEbr}
\end{center}
\end{table}

For the \ks\ lifetime we use the average $\tau_{\ks}= 0.08958\pm 0.00006$ ns \cite{PDG} obtained without assuming $CPT$ invariance.

\subsection{Two-pion modes}
The $\pi\pi (\gamma)$ terms are the largest.
  The value of $\alpha_{\pi\pi}$ of eq.(\ref{eq:alfa2pi}) is evaluated
  as:
\[
  \alpha_{\pi\pi} = \left(\frac{ \tau_{\ks}}{\tau_{\kl}}\x \BR(\kl\to \pi \pi)\x \BR(\ks\to \pi \pi) 
\right)^{1/2}   e^{i \phi_{\pi\pi}} \, . 
\]
 The values of  \BR(\ks\to\pic) and \BR(\ks\to\po\po), given in \Tab{tab:inputdata}, are determined
 from $\Rs$ using the constraint
 $\BR(\ks\to\pic) +  \BR(\ks\to\pio) + \BR(\ks\to \pi\ell\nu) = 1$;
  the value of $\BR(\ks\to \pi\ell\nu)$ is determined from the
  KLOE measurement of $\BR(\ks\to \pi e \nu)$~\cite{KLOE:BKSpienu} assuming lepton universality.
 For \BR$(\kl\to \pic)$ and \BR$(\kl\to \pio)$, we use the values from the KLOE average, which are  in  agreement  with recent measurements  from  KTeV~\cite{KTEV:BrL}.
  Finally, the values of $\phi_{+-}$ and $\phi_{00}$, the phases of $\eta_{+-}$
  and $\eta_{00}$, are taken from the PDG fit~\cite{PDG} without assuming $CPT$ invariance.

\begin{figure}[h!]
\begin{center}
    \resizebox{0.8\textwidth}{!}{\includegraphics{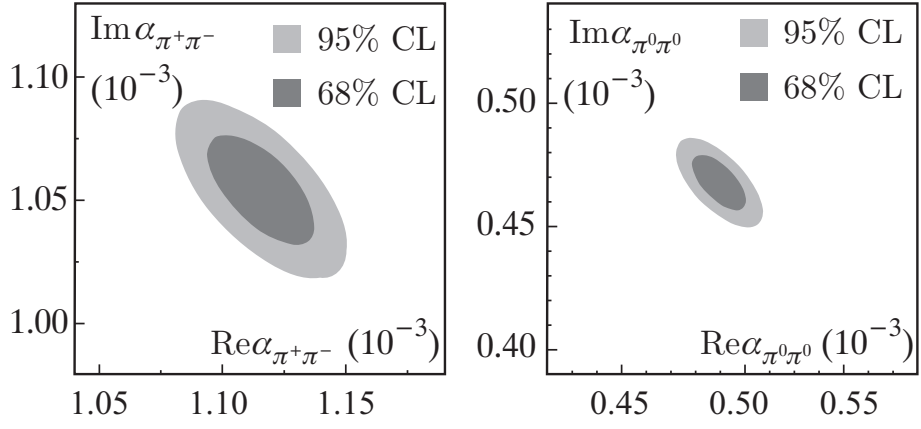}}
  \end{center}
  \caption{Representation of $\alpha_{\pic}$ and $\alpha_{\pio}$ in the complex plane. 
In each case, the two ellipses represent the 68\% and the 95\% CL contours.}
  \label{fig:alfa1}
\end{figure}

  Figure \ref{fig:alfa1} shows the 68\% and the 95\% CL contours 
  for $\alpha_{\pic}$ and $\alpha_{\pio}$. We find:
\begin{eqnarray}
 \alpha_{\pic} & = & \left(\vv(1.115 \pm 0.015) + i(1.055 \pm 0.015)\right)\x 10^{-3}\, , \nonumber\\ 
 \alpha_{\pio} & = & \left(\vv(0.489 \pm 0.007) + i(0.468 \pm 0.007)\right)\x 10^{-3}\, .\nonumber
\end{eqnarray}

\begin{figure}[h!]
\begin{center}
    \resizebox{0.8\textwidth}{!}{\includegraphics{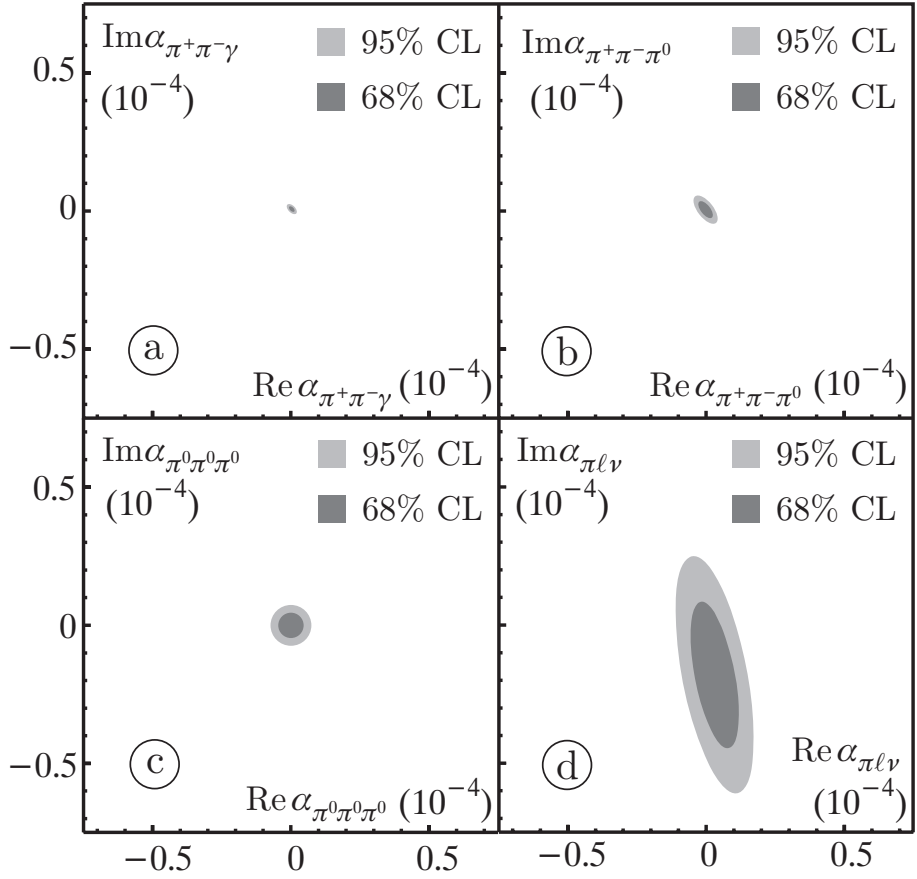}}
  \end{center}
  \caption{Bounds of the $\alpha$ values for three-body decays of \ks, \kl. 
Note that the same scale is used for all plots.}
  \label{fig:alfa2}
\end{figure}

 Following the discussion in \Sec{sec:ppg}, we determine $\alpha_{\pic\gamma_{\rm DE}}$
 from the measurement of $\eta_{+-\gamma}$ from Ref.~\cite{etappg} and 
 \BR$(\ks\to \pic\gamma)$ from the measurement of $\Rsg = \BR(\ks\to \pic\gamma)/\BR(\ks\to \pic)$~
\cite{BRKSppg}. The photon energy threshold was
 20 MeV for both measurements.
 The value of $\Rsg^{\rm th-IB}=
\left(\vv\BR(\ks\to\pi\pi\gamma)/\BR(\ks\to\pi\pi)\right)^{\rm th-IB} =\pt(0.700 \pm 0.001),-2,$
 is calculated  using the KLOE MC generator, which is described in Ref.~\citen{gattirad}.
 Figure \ref{fig:alfa2}(a) shows the 68\% and the 95\% CL contours 
 for $\alpha_{\pic\gamma_{DE}}$.
 We find a very small contribution:
$$
 \alpha_{\pic\gamma_{DE}}  = \left(\vv(5\pm7 ) + i(6\pm7)\right)\x 10^{-7}\, . 
$$

\subsection{Three-pion modes}
\label{sec:3pi}
     Until the recent results from  KLOE~\cite{KLOE:BKS3pi0} and NA48~\cite{NA48:BRS3pi0} became
available, the limiting
     contribution to the $CPT$ test was due to the $3\po$ final state. 
   The upper limit on $|\alpha_{\pio\po}|$ is  evaluated from \eq{eq:a000}
     using the KLOE  upper limit \BR(\ks\to
    3\po)$<\pt1.5,-7,$ at 95\% CL~\cite{KLOE:BKS3pi0} and the value of  \BR(\kl\to
    3\po) from the KLOE average. 
    The phase $\phi_{000}$ is taken as uniform in $\{0,2\pi\}$.
    For $\alpha_{\pic\po}$, \eq{eq:alfa3pi}, we use $\Re(\eta_{+-0})$ and $\Im(\eta_{+-0})$ as 
measured by CPLEAR~\cite{CPLEAR} and the value of \BR(\kl \to \pic\po) from the KLOE average, all given 
   in \Tab{tab:inputdata}.

  The 68\% and the 95\% CL contours for $\alpha_{\pic\po}$,
  and the 95\% CL contour for $\alpha_{\pio\po}$ are shown in 
  figures \ref{fig:alfa2}(b) and \ref{fig:alfa2}(c), respectively.
  We find:
\begin{eqnarray}
 \alpha_{\pic\po} & = &\left(\vv(0\pm 2) + i(0\pm2)\right)\x10^{-6}, \nonumber\\
 |\alpha_{\pio\po}| & < & 7\times 10^{-6} {\rm ~~at~95\% ~CL}\, .\nonumber
\end{eqnarray}

\subsection{Semileptonic modes}
\label{sec:semi}
     For the determination of $\alpha_{\pi\ell\nu}$ from \eq{eq:alfasl}, we combine 
     the KLOE measurement~\cite{KLOE:BKSpienu} of the \ks\ semileptonic charge
     asymmetry $A_S$ for \ks~ semileptonic decays, the PDG average~\cite{PDG} for 
    \kl~ semileptonic charge asymmetry $A_L$, and the CPLEAR time-dependent 
    measurement of \ko and \kob~  semileptonic rates~\cite{CPLEAR:semi}.
    The original CPLEAR result is given in \Tab{tab:cplear}.
\begin{table}[ht]
\begin{center}
\renewcommand{\arraystretch}{1.1} 
\begin{tabular}{@{}l|c|cccc}
               & value &  \multicolumn{4}{|c}{Correlation coefficients} \\
 \hline
 $\Re(\delta)$ & \pt( 3.0 \pm 3.4),-4, &   1   &       &       &       \\
$\Im(\delta)$  & \pt(-1.5 \pm 2.3),-2, & 0.44  &     1 &       &       \\ 
 $\Re(x_-)$    & \pt( 0.2 \pm 1.3),-2, &-0.56  & -0.97 &   1   &       \\
 $\Im(x_+)$    & \pt( 1.2 \pm 2.2),-2, &-0.60  & -0.91 &  0.96 &  1
\end{tabular}\\[2pt]
\caption{Values, errors, and correlation coefficients  for 
$\Re(\delta)$, $\Im(\delta)$, $\Re(x_-)$, and $\Im(x_+)$ 
 measured by CPLEAR. }
\label{tab:cplear}
\end{center}
\end{table}

  We have improved this result by adding the measurement 
  of $A_S-A_L=4[\Re(\delta)+\Re(x_-)]= (-2 \pm 10)\times10^{-3}$ \cite{KLOE:BKSpienu,PDG}. The
results, referred to as the $K_{\ell3}$ average, are given in \Tab{tab:kl3av}.
\begin{table}[ht]
\begin{center}
\renewcommand{\arraystretch}{1.1} 
\begin{tabular}{@{}l|c|ccccc}
               & value &  \multicolumn{5}{|c}{Correlation coefficients} \\
 \hline
 $\Re(\delta)$ & \pt( 3.4 \pm 2.8),-4,  &   1   &       &       &      &    \\
$\Im(\delta)$  & \pt(-1.0 \pm 0.7),-2,  & -0.27 &     1 &       &      &   \\ 
 $\Re(x_-)$    & \pt( -0.07 \pm 0.25),-2,&-0.23  & -0.58 &   1   &      &   \\
 $\Im(x_+)$    & \pt( 0.8 \pm 0.7),-2,  &-0.35  & -0.12 &  0.57 &  1   &   \\
 $A_S+A_L$     & \pt( 0.5 \pm 1.0),-2,  &-0.12  & -0.62 &  0.99 & 0.54 & 1   
\end{tabular}\\[2pt]
\caption{Values, errors, and correlation coefficients  for 
$\Re(\delta)$, $\Im(\delta)$, $\Re(x_-)$, $\Im(x_+)$, and $A_S+A_L$  
 obtained from a combined fit ($K_{\ell3}$ average). }
\label{tab:kl3av}
\end{center}
\end{table}
Finally, for $\BR(\kl \to \pi\ell\nu)$ we use the sum of $K_{e3}$ 
and $K_{\mu 3}$ branching ratios from the KLOE average.
Figure \ref{fig:alfa2}(d) shows the 68\% and the 95\% CL contours for $\alpha_{\pi\ell\nu}$. We find:
$$
 \alpha_{\pi\ell\nu} = \left(\vv(0.3\pm0.6) + i(-1.8\pm1.8)\right)\x10^{-5}\, .
$$

\section{Results}
\label{sec:resu}    

  Inserting  the values of the $\alpha$ parameters 
  into \eq{eq:b-s_sol}, we obtain\footnote{ The accuracy on $\Re(\epsilon)$
  improves by about 30\% with inclusion of the KTeV measurement of $\BR(\kl\to \pic)$ \cite{KTEV:BrL}.} :
\begin{equation} \Re(\epsilon) = (159.6 \pm 1.3)\times 10^{-5} , \qquad   \Im(\delta) = (0.4 \pm 2.1)\times  10^{-5}\, ,
\label{klores}
\end{equation}
 where  all correlations  among the input data are taken into account, including
that from the direct determination of $\Im(\delta)$ from the semileptonic decays. 
 The complete information is given in \Tab{tab:fine}.

\begin{table}[hbt]
\begin{center}
\renewcommand{\arraystretch}{1.1} 
\begin{tabular}{@{}l|c|cccc}
               & value &  \multicolumn{4}{|c}{Correlation coefficients} \\
 \hline
$\Re(\epsilon)$ & \pt( 159.6 \pm 1.3),-5, &   1    &       &       &       \\
 $\Im(\delta)$  & \pt( 0.4 \pm 2.1),-5,   & -0.17  &     1 &       &       \\ 
 $\Re(\delta)$  & \pt( 2.3 \pm 2.7),-4,   &  0.20  & -0.22 &   1   &       \\
 $\Re(x_-)$     & \pt( -2.9 \pm 2.0),-3,  & -0.25  &  0.37 & -0.49 &  1
\end{tabular}\\[2pt]
\caption{Summary of results: values, errors, and correlation coefficients
 for $\Re(\epsilon)$,   $\Im(\delta)$, $\Re(\delta)$, and $\Re(x_-)$.}
\label{tab:fine}
\end{center}
\end{table}

\begin{figure}[h!]
  \begin{center}
    \resizebox{0.8\textwidth}{!}{\includegraphics{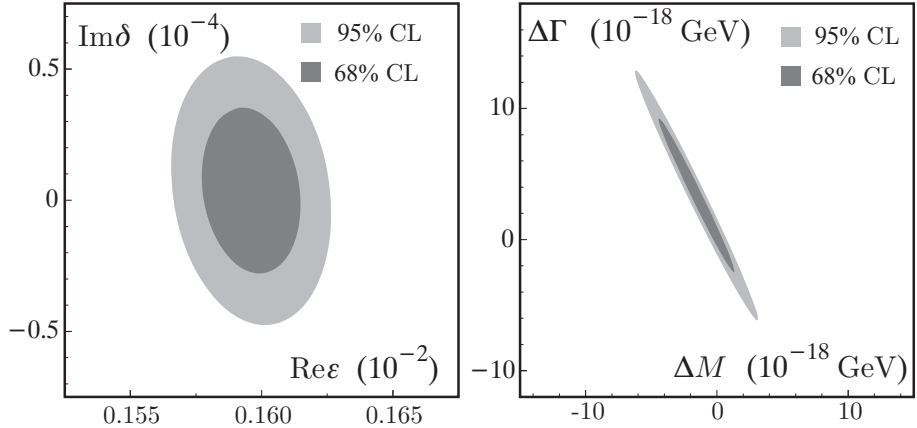}}
  \end{center}
  \caption{Left: allowed region at 68\% and 95\% CL in the $\Re(\epsilon)$, $\Im(\delta)$ plane.
Right: allowed region at 68\% and 95\% CL in the 
$\Delta M, \Delta \Gamma$ plane.}
  \label{fig:final}
\end{figure}

The allowed region in the $\Re(\epsilon)$, $\Im(\delta)$ plane
   at 68\% CL and 95\% CL is shown in the left panel of \Fig{fig:final}.
   A small correlation between $\Re(\epsilon)$ and $\Im(\delta)$ is evident, which is due to the
   semileptonic term. With the new KLOE data used in the present analysis, the process
   giving the largest contribution to the size of the allowed region is now 
$\kl \to \pic$, through the uncertainty on $\phi_{+-}$.
   Besides the $\pi\pi$ final states, only the semileptonic term gives an appreciable 
   contribution ($\sim 10\%$) to the error on $\Im(\delta)$.   
Our results, Eq. \ref{klores}, are a significant improvement over those of CPLEAR~\cite{CPLEAR:bs}:
  $$ \Re(\epsilon) = (164.9 \pm 2.5)\times 10^{-5},\qquad   \Im(\delta) = (2.4 \pm 5.0)\times
  10^{-5}. $$
Note also that the central value of $ \Re(\epsilon)$ is quite different. This is due to the new measurement of \BR(\kl\to\pic)~\cite{KLOE:bpp}.
 
    The limits on $\Im(\delta)$ and $\Re(\delta)$ can be used
   to constrain the mass and width difference between $\ko$
   and $\kob$ via
  $$ \delta = { { i(m_{\ko} - m_{\kob}) + {1\over2} (\Gno-\Gamma_{\kob}) }  
       \over{\GAMS-\GAML} }\cos\phi_{SW} e ^{i \phi_{SW}} [1+{\mathcal O}(\epsilon)]\, .
     $$
The allowed region in the $\Delta M = (m_{\ko} - m_{\kob}),
\Delta \Gamma = (\Gno-\Gamma_{\kob}) $ plane is shown in the right panel of \Fig{fig:final}. 
The strong correlation reflects the high precision of $\Im(\delta)$
compared to $\Re(\delta)$. Since the total decay widths are dominated by long-distance 
dynamics, in models where $CPT$ invariance is a pure short-distance phenomenon,
it is useful to consider the limit $\Gno=\Gamma_{\kob}$. In this 
limit (i.e. neglecting $CPT$-violating effects in the decay amplitudes),
we obtain the following bound on the neutral kaon mass difference:

  $$ -5.3  \times 10^{-19}~ {\rm GeV}< m_{\ko} - m_{\kob}  < 6.3 \times 10^{-19}~ \rm{GeV \quad at~ 95 ~\% ~CL}\, . $$
Our result represents a significant improvement with respect to 
$|m_{\ko} - m_{\kob}|<12.7 \times 10^{-19}~{\rm GeV}$ at 90\%~CL, obtained by
CPLEAR~\cite{CPLEAR}. 

\section*{Acknowledgments}

This work was supported in part by DOE grant DE-FG-02-97ER41027; 
by EURODAPHNE, contract FMRX-CT98-0169; 
by the German Federal Ministry of Education and Research (BMBF) contract 06-KA-957; 
by Graduiertenkolleg `H.E. Phys. and Part. Astrophys.' of Deutsche Forschungsgemeinschaft,
Contract No. GK 742; by INTAS, contracts 96-624, 99-37; 
by TARI, contract HPRI-CT-1999-00088.

\appendix 
\section{The $\pi\pi\gamma$ contribution to the $\alpha_i$}
\label{sec:appendix}
By construction, the leading contribution of the $\pi\pi\gamma$
state to the unitarity sum, namely the interference of the \kl\ and \ks\ 
bremsstrahlung amplitudes, is included in $\alpha_{+-(\gamma)}$.
The largest sub-leading term missing in $\alpha_{+-(\gamma)}$
is the DE-brems\-stra\-hlung interference, which we include in $\alpha_{\pi\pi\gamma_{\rm E1-S(L)}}$  of \eq{eq:alfappg1}.
To evaluate this contribution, we introduce the total (IB+DE) amplitude ratio
\beq 
\eta_{+-\gamma}(E_\gamma) = \frac{ \cA_L(\pi\pi\gamma_{\rm IB+E1}) }{ 
\cA_S(\pi\pi\gamma_{\rm IB+E1}) }
= \eta_{+-} + \epsilon^{\,\prime}_{+-\gamma}(E_\gamma)
\label{eq:etappg1}
\eeq
This ratio is an observable quantity which can be measured in an
interference experiment~\cite{etappg}. As explicitly indicated, 
$\eta_{+-\gamma}$ depends on the photon energy and 
it can be decomposed into the energy-independent 
parameter $\eta_{+-}$ and the direct-$CP$-violating
term $\epsilon^{\,\prime}_{+-\gamma}$. 
In the $E_\gamma\to 0$ limit  $\epsilon^{\,\prime}_{+-\gamma} \propto E_\gamma^2$  
\cite{DI}.
Using \eq{eq:etappg1}, we can write 
\beqa
&& \frac{1}{\Gamma_S} \left\langle\vv\cA_L(\pi\pi\gamma)  \cA^*_S(\pi\pi\gamma_{\rm E1})
+\cA_L(\pi\pi\gamma_{\rm E1})\cA^*_S(\pi\pi\gamma)\right\rangle=\no \\ 
&&  = \frac{1}{\Gamma_S}
    \left\langle\vv\:\eta_{+-}\:\left(\vv\cA_S(\pi\pi\gamma)  \cA^*_S(\pi\pi\gamma_{\rm E1})
    +\cA_S(\pi\pi\gamma_{\rm E1})   \cA^*_S(\pi\pi\gamma) \right) + \right. \no \\
&& \qquad  \quad \left. + \epsilon^{\,\prime}_{+-\gamma}~  \cA_S(\pi\pi\gamma)
\cA^*_S(\pi\pi\gamma) (1+ \delta_{\rm E1})\right\rangle \no \\
&& \approx  ~\eta_{+-}~\Delta B(\ks\to\pi\pi\gamma_{\rm DE}) + 
 \frac{1}{\Gamma_S}\int dE_\gamma~\epsilon^{\,\prime}_{+-\gamma}~ \frac{d \Gamma(\ks \to \pi\pi\gamma)}{d E_\gamma }
\label{eq:alfappg3}
\eeqa
where $\delta_{\rm E1} = \cA_S(\pi\pi\gamma_{\rm E1})/ \cA_S(\pi\pi\gamma)$ 
is a very small overall correction factor which can be safely neglected and 
$\Delta B(\ks\to\pi\pi\gamma_{\rm DE}) = \BR(\ks\to\pi\pi\gamma)^{\rm exp} - 
\BR(\ks\to\pi\pi\gamma)^{\rm th-IB}$ is the deviation of the observed 
$\ks\to\pi\pi\gamma$ decay rate from that inferred from a 
pure bremsstrahlung spectrum. By construction, the integral on the right-hand 
side of \eq{eq:alfappg3} is infrared safe. Since at present 
there is no evidence for a non-vanishing $\epsilon{\,^\prime}_{+-\gamma}$ \cite{etappg},
in  \eq{eq:alfappg2} we replace this integral with the product 
$(\eta_{+-\gamma}-\eta_{+-}) \times \BR(\ks \to \pi\pi\gamma)$, where 
$\BR(\ks \to \pi\pi\gamma)$ indicates the branching fraction for 
a real photon emission with minimum photon-energy cut  
equivalent to that used in the corresponding $\eta_{+-\gamma}$  measurement
($E^{\rm cut}_\gamma=$ 20 MeV in Ref.~\citen{etappg}). The contribution to \eq{eq:alfappg3} 
generated by $K_{L,S} \to \pi\pi\gamma$ amplitudes  
with $E_\gamma < E^{\rm cut}_\gamma$ 
vanishes in the limit $E_\gamma^{\rm cut} \to 0$  and can be safely neglected.

\end{document}